\documentclass[conference]{IEEEtran}
\IEEEoverridecommandlockouts
\usepackage{cite}
\usepackage{amsmath,amssymb,amsfonts}
\usepackage{graphicx}
\usepackage{textcomp}
\usepackage{amsmath,amssymb,amsfonts}
\usepackage{booktabs}
\usepackage{graphicx}
\usepackage{textcomp}
\usepackage{comment}
\usepackage{float}
\usepackage{verbatim}
\usepackage{caption}
\usepackage{pbalance}
\usepackage[caption=false,font=normalsize,labelfont=sf,textfont=sf]{subfig}
\usepackage{multirow}
\usepackage{booktabs}
\usepackage{hhline}
\usepackage{tikz}
\usepackage{mdframed}
\usepackage[caption=false,font=normalsize,labelfont=sf,textfont=sf]{subfig}
\usepackage{caption}
\usepackage[utf8]{inputenc}
\usepackage{algorithm}
\usepackage{placeins}
\usepackage{placeins}
\usepackage{pbalance}
\usepackage{algpseudocode}
\usepackage{xcolor}
\usepackage{algorithm}
\def\BibTeX{{\rm B\kern-.05em{\sc i\kern-.025em b}\kern-.08em
    T\kern-.1667em\lower.7ex\hbox{E}\kern-.125emX}}
\begin{document}


\title{Improved Brain Tumor Detection in MRI: Fuzzy Sigmoid Convolution in Deep Learning\\
{\footnotesize \textsuperscript{*}Note: Accepted in IEEE IJCNN 2025}
\thanks{This work is supported by the Finnish National Agency for Education (Grant No OPH-2023), and FCFH Support Funding for 2024—Grants. 
}
}
 
\author{
\IEEEauthorblockN{
Muhammad Irfan\IEEEauthorrefmark{1}\IEEEauthorrefmark{2}
Anum Nawaz\IEEEauthorrefmark{3},
Riku Kl\'en\IEEEauthorrefmark{1},
Abdulhamit Subasi\IEEEauthorrefmark{4},
Tomi Westerlund\IEEEauthorrefmark{2}
and 
Wei Chen\IEEEauthorrefmark{5}
}
\IEEEauthorblockA{
\IEEEauthorrefmark{1}Turku PET Center, Turku University Hospital, University of Turku, Finland 
\\
\IEEEauthorrefmark{2}Turku Intelligent Embedded and Robotic Systems lab, University of Turku, Finland
\\
\IEEEauthorrefmark{3}School of Information Science and Technology, Fudan University, China
\\
\IEEEauthorrefmark{4}Information Sciences and Technology, University at Albany, USA
\\
\IEEEauthorrefmark{5}University of Sydney, Australia}
\{muhammad.m.irfan,
anum.nawaz,
tomi.westerlund, 
riku.klen\}@utu.fi\\
asubasi@albany.edu, 
wei.chenbme@sydney.edu.au
}

\maketitle

\begin{abstract}
Early detection and accurate diagnosis are essential to improving patient outcomes. The use of convolutional neural networks (CNNs) for tumor detection has shown promise, but existing models often suffer from overparameterization, which limits their performance gains. In this study, fuzzy sigmoid convolution (FSC) is introduced along with two additional modules: top-of-the-funnel and middle-of-the-funnel. The proposed methodology significantly reduces the number of trainable parameters without compromising classification accuracy. A novel convolutional operator is central to this approach, effectively dilating the receptive field while preserving input data integrity. This enables efficient feature map reduction and enhances the model's tumor detection capability. In the FSC-based model, fuzzy sigmoid activation functions are incorporated within convolutional layers to improve feature extraction and classification. The inclusion of fuzzy logic into the architecture improves its adaptability and robustness. Extensive experiments on three benchmark datasets demonstrate the superior performance and efficiency of the proposed model. The FSC-based architecture achieved classification accuracies of 99.17\,\%, 99.75\,\%, and 99.89\,\% on three different datasets. The model employs 100 times fewer parameters than large-scale transfer learning architectures, highlighting its computational efficiency and suitability for detecting brain tumors early. This research offers lightweight, high-performance deep-learning models for medical imaging applications.

\end{abstract}


\section{Introduction}
\label{intro}

The human brain functions as the central hub for many vital functions such as memory, speech, cognition, and motor skills~\cite{i1,10371274,10842226, 10950432, 10918671}. An irregular brain tissue growth can disrupt these functions, leading to brain tumors~\cite{i2}. Brain tumors are a major concern due to their various types and potential severity, presenting a significant challenge to both patients and healthcare providers. In the United States alone, over a million individuals are affected by brain tumors, with over 94,000 new cases projected in the coming year. Among these, approximately 72\,\% are benign, while 28\,\% are malignant, impacting people across various age groups~\cite{10892686}.

In children aged 0–14, brain tumors are the most common solid cancer diagnosis, with an 81.3\,\% five-year survival rate. They also rank as the second most frequent cancer among adolescents and young adults aged 15–39, where the five-year relative survival rate stands at 90.9\,\%. For individuals aged 40 and above, who represent the majority of cases, the survival rate is 72.5\,\% for malignant tumors and 90.3\,\% for non-malignant ones. It is clear from these statistics that more research and innovative solutions are needed to improve diagnosis, treatment, and quality of life for brain tumor patients across all ages~\cite{10892686}.

Magnetic resonance imaging (MRI) is an important part of brain tumor diagnosis and treatment due to its non-invasive nature and high sensitivity, which avoids ionizing radiation~\cite{i3}. However, it relies heavily on expert knowledge for interpretation. In recent years, intelligent systems, particularly deep learning algorithms using MRI data, have gained significant attention for their potential in brain tumor detection~\cite{i4,10043011}. Deep Convolutional Neural Networks (DCNNs), a subset of artificial neural networks, have shown promising results in medical imaging and diagnostic applications~\cite{shahid2021deep,8743438}.  This research evaluates the accuracy of the proposed approach for brain tumor classification using only 0.21627 million parameters.

\subsection*{Main Contributions of the Paper}

\begin{enumerate}


\item FuzzySigmoidConv (FSC): This framework directly integrates fuzzy logic into convolutional operations. The architecture comprises three modules: TOFU (Top of Funnel), MOFU (middle of Funnel), and FSC (FuzzySigmoidConv). With TOFU, features are aggregated progressively by applying dilation and fuzzy membership-based optimization. By computing \textit{high} and \textit{low} membership functions using sigmoid activations, this module introduces a dynamic feature weighting mechanism. The FSC-based model consists of 216,270 trainable parameters, providing a more complex feature extraction process optimized for nuanced tasks like brain tumor classification.

\item {Performance and Efficiency Trade-offs:} The FSC architecture demonstrates exceptional classification accuracy for brain tumor detection while maintaining resource efficiency through its reduced parameter count.

\item Datasets: The proposed method showed exceptional accuracy, F1-score, and Kappa performance on three datasets. This evidence indicates that fuzzy logic can help improve feature extraction. 

\end{enumerate}

The paper is organized as follows: \textbf{Section~\ref{sec:related_work}} summarizes existing research on brain tumor classification. \textbf{Section~\ref{sec:Methodology}} outlines the datasets used, preprocessing techniques applied, and the implementation details of the FSC modules, including TOFU and MOFU components. \textbf{Section~\ref{sec:results}} evaluates the proposed approach, comparing its performance to state-of-the-art transfer learning models, and analyzing various experimental scenarios. Finally, \textbf{Section~\ref{sec:conclusions}} concludes the paper by detailing the main findings and discussing the broader implications of this work.

\section{Related Work}
\label{sec:related_work}

In recent years, deep learning techniques have gained significant attention for detecting and classifying brain tumors based on MRI data~\cite{10892686,10361450}. Various architectures and strategies have been investigated, contributing to advancements in tumor diagnosis~\cite{chaki2023deep,8743438}. There remains a high level of trust in convolutional neural networks (CNNs) for their ability to improve classification accuracy in tasks ranging from binary to multi-class tumor classification. The methods continue to advance, with novel features and approaches being developed to address the challenges associated with precision and reliability in clinical applications.

In an ensemble learning approach, Kang et al. \cite{kang2021mri} combined deep feature extraction with ResNeXt-101 and SVM classifiers, resulting in high test accuracy. Agarwal et al. \cite{tablecomp2} applied the VGG16 model for brain tumor classification, achieving 90\,\% accuracy. Yentür et al. \cite{tablecomp1} employed kernel SVM for brain tumor classification using MRI images, achieving 97\,\% accuracy. Çınar et al. \cite{r12} evaluated deep learning models such as ResNet101, VGG19, InceptionV3, DenseNet, and AlexNet, with test accuracies ranging from 89.5\,\% to 98.6\,\%. Tang et al. \cite{tang2023gam} introduced GAM-SpCaNet, a spinal convolution attention network, and achieved an accuracy of 99.28\,\% on the BRATS 2019 dataset. Kesav and Jibukumar \cite{kesav2022efficient} proposed a low-complexity brain tumor identification model based on RCNN and CNN, showing 98.21\,\% accuracy but requiring a longer execution time. Anagun \cite{anagun2023smart} developed a CNN-based diagnosis scheme using EfficientNetv2, achieving 99.85\,\% accuracy. Sathish and Elango \cite{sathish2022gaussian} proposed a hybrid fuzzy clustering approach with a radial basis neural network, with 79.98\,\% accuracy. Despite advancements, a higher error rate was noted in their approach. In \cite{mohanty2024feature}, the authors employed a Soft Attention-based CNN framework to classify Meningioma, Glioma, and Pituitary tumors, achieving an overall accuracy of 95.1\%. The model attained class-specific accuracies of 97.97\% for Meningioma, 97.1\% for Glioma, and 97.79\% for Pituitary tumors, outperforming other class-specific approaches.

Nagarani et al.~\cite{nagarani2024self} presented a novel approach that produced outstanding results, with an F1-score of 99.77\,\%, alongside an impressively low error rate. However, the model's computational demands were substantial, requiring 25.5 million parameters, which makes it unsuitable for environments with limited resources. Deepa et al. \cite{deepa2023hybrid} presented a hybrid optimization algorithm for brain tumor segmentation and classification using MRI data. The model extracts and feeds CNN features into a deep residual network (DRN) trained using the Chronological Jaya Honey Badger Algorithm (CJHBA). The model achieved 92.10\,\% accuracy, 93.13\,\% sensitivity, and 92.84\,\% specificity on the BRATS 2018 dataset. Anaya-Isaza et al. \cite{anaya2023optimizing} developed a framework to explore state-of-the-art deep learning architectures for brain tumor classification and detection. They evaluated popular networks such as InceptionResNetV2, DenseNet121, and ResNet50V2, achieving over 97\,\% accuracy on the Figshare dataset, focusing on classifying glioma, meningioma, and pituitary tumors. Dahiwade et al. \cite{dahiwade2019designing} proposed a machine learning-based disease prediction model using KNN and CNN algorithms with an accuracy of 84.5\,\%.

Raju and Rao \cite{raju2023classification} developed a hybrid optimization-based deep learning model for brain tumor detection. The process began with pre-processing using an adaptive Wiener filter, followed by segmentation using U-Net. The model applied data augmentation to improve image quality and extracted features for classification. DenseNet, trained with Red Deer Tasmanian Devil Optimization (RDTDO), was used for first-level classification, followed by a second-level classification using a DRN. The model achieved a maximum accuracy of 94.7\,\%. Saidani et al. \cite{saidani2023enhancing} introduced a two-phase approach for brain tumor classification that utilized both image-based and data-based features. The first phase employed a UNet transfer learning model for classifying patients as either having a brain tumor or being normal. The second phase combined 13 features with a voting classifier that integrated deep convolutional layers with stochastic gradient descent and logistic regression, resulting in an accuracy of 99\,\%. Li et al. \cite{li2023mri} proposed a two-stage deep learning model for the automatic detection and segmentation of brain metastases (BMs) in MRI images. The model consists of a lightweight segmentation network and a multi-scale classification network for false-positive suppression. On a dataset of 649 patients, the model achieved a sensitivity of 90\,\%, precision of 56\,\%, and an average dice score of 81\,\%, outperforming one-stage models, particularly for smaller BMs (under 5 mm), with a sensitivity of 66\,\%. Ansari et al. \cite{ansari2023numerical} proposed an enhanced methodology to employ fuzzy clustering for tumor segmentation, followed by feature extraction and improved Support Vector Machine for classification, achieving an overall accuracy of 88\%.
The studies highlighted in Section~\ref{sec:related_work} demonstrate the multitude of methods and models developed for brain tumor classification. While high-performing models such as SPGAN-MSOA-CBT-MRI~\cite{nagarani2024self} achieve remarkable precision, their reliance on millions of parameters imposes significant computational overhead. To overcome these challenges, this paper introduces an efficient approach—FSC—incorporating two specialized modules, TOFU and MOFU. Using MRI data, this method delivers highly accurate brain tumor classifications with only 216.27K parameters, significantly reducing computational time.

\section{Materials and Methods}
\label{sec:Methodology}

In this section, we present detailed insights of the deployed datasets to highlight the methodological classification and characteristics, facilitating replication and understanding of the adopted methods. Furthermore,  we provide the insights of proposed convolution architecture and its implementation in the proposed FSC network architecture.

\subsection{Datasets for Brain Tumor Classification}

\subsubsection{Dataset I}

The dataset I, sourced from Kaggle\textquotesingle{}s ``Brain MRI data for Diagnosing Brain Tumor" collection~\cite{r2}, serves a crucial role in our proposed system by providing the baseline to explore binary classification problems related to brain tumor detection while exploiting magnetic resonance images.  The dataset comprises a total of 3,060 2D images, with 1,500 indicating tumor presence and the same number depicting non-tumor instances. Unlabelled 60 images have been excluded from our MRI scan analysis.

To ensure a robust experimental framework, the dataset was partitioned into training and testing subsets. Specifically, the training data was stratified further, setting 70\,\% for model training and the remaining 30\,\% for validation purposes and testing. The images were systematically organized into different folders and titled accordingly: tumor images were prefixed with “y” (y1, y2, and y3 ), while non-tumor images were prefixed with “no” (no1, no2, no3 ). Representative samples from the dataset are illustrated in Figure 1. Figure 1a and Figure 1c are drawn from dataset I, Figure 1a depicts an image without a tumor while Figure 1c shows tumor examples.



\subsubsection{Dataset II}

\begin{figure}[th]
\centering
\subfloat[]{\includegraphics[width=1.0in, height=1.0in]{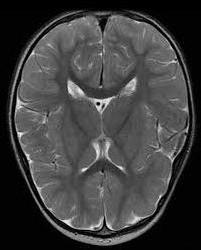}
\label{fig_a}}
\subfloat[]{\includegraphics[width=1.0in, height=1.0in]{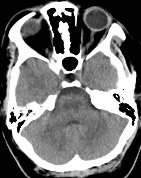}
\label{fig_b}}
\hfill
\subfloat[]{\includegraphics[width=1.0in, height=1.0in]{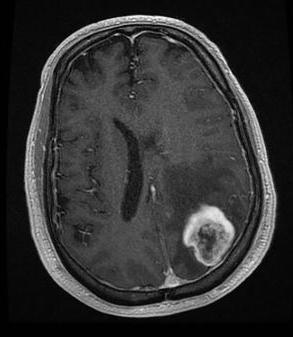}
\label{fig_c}}
\subfloat[]{\includegraphics[width=1.0in, height=1.0in]{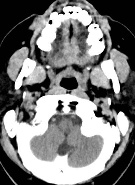}
\label{fig_d}}
\caption{Samples images from the datasets. (a) and (b) shows
no tumor samples while (c) and (d) show with tumor.}
 \label{fig:ref}
\end{figure}

In the course of a comprehensive analysis, dataset II~\cite{dataset2}, originally developed for segmentation tasks and containing image masks, was modified to align with the requirements of classification tasks by excluding the masks.  The dataset comprises a total number of 2501 images, of which 950 exhibit tumors and 1551 without tumor. As such, the dataset demonstrates an alternative platform for evaluating the classification model. Figure~\ref{fig_b} and Figure~\ref{fig_d} are obtained from dataset II, Figure~\ref{fig_b} shows an image without a tumor while Figure~\ref{fig_d} represents samples with a tumor.

The dataset was partitioned into approximately 70\,\% for training and 30\,\% for validation and testing. The systematic method enabled the archiving of images in different folders, providing a coherent naming convention, wherein tumor and non-tumor images were prefixed with “y” and “no” respectively.

\subsubsection{Dataset III}

Dataset III was constructed by combining datasets I and II, which increases data versatility and reliability. This aggregated dataset comprises of 5,501 images, of which 2,450 show tumors, while the remaining ones 3,051 are without tumors. Combining the datasets allows for a more detailed analysis of classification problems by giving more expanded and vast collection of data.

In a systematic partitioning scheme, approximately 70\,\% of the dataset was reserved for training, while 30\,\% was aligned for validation and testing. This method ensured that our results were reliable due to a balanced distribution. Moreover, we ensured that images from both Dataset I and Dataset II shown in the training, validation, and testing sets. More concisely we divided the data in an equal number of MRI images from both datasets I and II. Test set contained 450 images from Dataset I and 450 images from Dataset II.  


\subsection{Preprocessing}

The proposed methodology employed a preprocessing timeline to prepare brain MRI images for model training, validation, and testing. We used same techniques to augment training data. The augmentation comprises of random horizontal and vertical flips to introduce variability in image orientation, random rotations up to twenty degrees to account for variations in scan angles, and color jitter to adjust brightness, contrast, saturation, and hue to simulate different conditions during MRI acquisition. To ensure consistency across datasets, all images were normalized using the mean and standard deviation values derived from ImageNet. In the proposed methodology, smaller images were used without padding to reduce computational complexity and memory constraints, balancing efficiency with GPU-based feature capture.





To accommodate varying memory constraints during experimentation, images were resized to fixed resolutions of either $128 \times 128$ or $62 \times 62$ pixels, depending on the specific experimental setup. Furthermore, smaller sizes were used to reduce computational load and memory requirements, which is crucial to efficient training when GPU resources are limited. The approach did not include the reflection padding step and relied solely on resizing after common augmentations. This straightforward pipeline allowed for faster processing, and when noise was added to the images, the smaller size helped achieve better results. The model focused on more prominent features, enhancing performance despite the smaller resolution of images.

\paragraph{\normalsize{\textbf{Validation and Test Data}}}

In this approach, the validation and test datasets were not augmented. These datasets were only resized to the appropriate dimensions ($128 \times 128$, or $62 \times 62$ pixels) and normalized using the same mean and standard deviation values as the training data. This ensured that the evaluation datasets remained consistent with the input used during training, allowing for a reliable assessment of model performance.

\subsection{Layer-by-layer development of Approach}
\label{sec:appII}

In this section, we introduce FSC, an innovative convolutional layer that integrates fuzzy logic principles with convolutional neural networks. FSC improves the feature extraction abilities of traditional convolution layers by dynamically adjusting the contributions of various feature activation\textquotesingle{}s through fuzzy membership functions based on sigmoid activation. The overall network structure is composed of three major components: TOFU blocks, MOFU blocks, and fully connected layers.

The FSC layer applies fuzzy membership functions to the convolution process, modulating each feature map through two sigmoid-based membership functions: \textit{high membership} and \textit{low membership}. These functions adjust the importance of individual features based on their activation values.

Given an input \( x \) passed through a convolution operation and batch normalization, the fuzzy membership functions are defined as follows:

\begin{equation}
    \text{high\_membership} = \sigma(x),
\end{equation}
\begin{equation}
    \text{low\_membership} = \sigma(-x),
\end{equation}
where \( \sigma(\cdot) \) is the sigmoid activation function.

The output of the FSC layer, denoted as \( o_{\text{fsc}} \), is computed by combining the feature map \( x \) with the high and low membership functions as follows:

\begin{equation}
    o_{\text{fsc}} = \text{high\_membership} \cdot x + \text{low\_membership} \cdot (1 - x).
\end{equation}

This mechanism allows the network to emphasize more critical features while suppressing less relevant ones dynamically. After computing \( o_{\text{fsc}} \), the result is passed through a ReLU activation function:

\begin{equation}
    o_{\text{relu}} = \text{ReLU}(o_{\text{fsc}}).
\end{equation}

The implementation of FSC is given in algorithm~\ref{alg:fsc}.

\begin{algorithm}
\caption{FuzzySigConv}
\label{alg:fsc}
\begin{algorithmic}[1]

\State \textbf{Class} \texttt{FuzzySigConv}
\State \textbf{Function} \texttt{\_\_init\_\_}(in\_channels, out\_channels, kernel\_size=3, dilation=2, padding=1)
\State \quad \texttt{self.conv} $\leftarrow$ \texttt{Conv2D}(in\_channels, out\_channels, kernel\_size, dilation=dilation, padding=padding)
\State \quad \texttt{self.bn} $\leftarrow$ \texttt{BatchNorm2D}(out\_channels)

\State \textbf{Function} \texttt{fuzzy\_membership}(x)
\State \quad \texttt{high\_membership} $\leftarrow$ \texttt{Sigmoid}(x)
\State \quad \texttt{low\_membership} $\leftarrow$ \texttt{Sigmoid}(-x)
\State \quad \textbf{Return} \texttt{high\_membership}, \texttt{low\_membership}

\State \textbf{Function} \texttt{forward}(x)
\State \quad x $\leftarrow$ \texttt{self.conv}(x)
\State \quad x $\leftarrow$ \texttt{self.bn}(x)
\State \quad \texttt{high\_membership}, \texttt{low\_membership} $\leftarrow$ \texttt{fuzzy\_membership}(x)
\State \quad x $\leftarrow$ \texttt{high\_membership} $\times$ x $+$ \texttt{low\_membership} $\times$ (1 - x)
\State \quad \textbf{Return} \texttt{ReLU}(x)

\end{algorithmic}
\end{algorithm}

\begin{algorithm}
\caption{TOFU: FSC}
\label{alg:tofu_fsc}
\begin{algorithmic}[1]

\State \textbf{Class} \texttt{TOFU}
\State \textbf{Function} \texttt{\_\_init\_\_}(in\_channels, out\_channels)
\State \quad \texttt{conv1} $\leftarrow$ \texttt{FuzzySigConv}(in\_channels, 16, kernel\_size=3, dilation=2, padding=2)
\State \quad \texttt{conv2} $\leftarrow$ \texttt{FuzzySigConv}(16, 32, kernel\_size=3, dilation=2, padding=2)
\State \quad \texttt{conv3} $\leftarrow$ \texttt{FuzzySigConv}(32 + 16, 48, kernel\_size=3, dilation=2, padding=2)
\State \quad \texttt{compress} $\leftarrow$ \texttt{Conv2D}(96, out\_channels, kernel\_size=1)
\State \quad \texttt{bn} $\leftarrow$ \texttt{BatchNorm2D}(out\_channels)

\State \textbf{Function} \texttt{forward}(inputs)
\State \quad x1 $\leftarrow$ \texttt{conv1}(inputs)
\State \quad x2 $\leftarrow$ \texttt{conv2}(x1)
\State \quad x3 $\leftarrow$ \texttt{conv3}(\texttt{Concat}(x1, x2, dim=1))
\State \quad x4 $\leftarrow$ \texttt{ReLU}(\texttt{compress}(\texttt{Concat}(x1, x2, x3, dim=1)))
\State \quad \textbf{Return} \texttt{bn}(x4)

\end{algorithmic}
\end{algorithm}

\begin{algorithm}[!ht]
\caption{MOFU: FSC}
\label{algo:mofu_fsc}
\begin{algorithmic}[1]

\State \textbf{Class} \texttt{MOFU}
\State \textbf{Function} \texttt{\_\_init\_\_}(in\_channels, out\_channels)
\State \quad \texttt{conv1} $\leftarrow$ \texttt{FuzzySigConv}(in\_channels, 16, kernel\_size=3, dilation=2, padding=2)
\State \quad \texttt{conv2} $\leftarrow$ \texttt{FuzzySigConv}(16, 32, kernel\_size=3, dilation=2, padding=2)
\State \quad \texttt{conv3} $\leftarrow$ \texttt{FuzzySigConv}(32, out\_channels, kernel\_size=3, dilation=2, padding=2)
\State \quad \texttt{bn} $\leftarrow$ \texttt{BatchNorm2D}(out\_channels)

\State \textbf{Function} \texttt{forward}(inputs)
\State \quad x $\leftarrow$ \texttt{conv1}(inputs)
\State \quad x $\leftarrow$ \texttt{conv2}(x)
\State \quad x $\leftarrow$ \texttt{conv3}(x)
\State \quad \textbf{Return} \texttt{bn}(x)

\end{algorithmic}
\end{algorithm}

\begin{algorithm}[!ht]
\caption{Net: FSC} 
\label{algo:fsc_net}
\begin{algorithmic}[1]

\State \textbf{Class} \texttt{Net}
\State \textbf{Function} \texttt{\_\_init\_\_}(num\_classes=1)
\State \quad \texttt{initial\_conv} $\leftarrow$ \texttt{Conv2D}(3, 32, kernel\_size=3, stride=1, padding=1)
\State \quad \texttt{initial\_bn} $\leftarrow$ \texttt{BatchNorm2D}(32)
\State \quad \texttt{tofu1} $\leftarrow$ \texttt{TOFU}(in\_channels=32, out\_channels=DIM)
\State \quad \texttt{tofu2} $\leftarrow$ \texttt{TOFU}(in\_channels=DIM, out\_channels=128)
\State \quad \texttt{mofu} $\leftarrow$ \texttt{MOFU}(128, 256)
\State \quad \texttt{global\_avg\_pool} $\leftarrow$ \texttt{AdaptiveAvgPool2d}(1)
\State \quad \texttt{fc1} $\leftarrow$ \texttt{Linear}(256, 256)
\State \quad \texttt{fc2} $\leftarrow$ \texttt{Linear}(256, 128)
\State \quad \texttt{classifier} $\leftarrow$ \texttt{Linear}(128, num\_classes)

\State \textbf{Function} \texttt{forward}(x)
\State \quad x $\leftarrow$ \texttt{ReLU}(\texttt{initial\_bn}(\texttt{initial\_conv}(x)))
\State \quad x $\leftarrow$ \texttt{tofu1}(x)
\State \quad x $\leftarrow$ \texttt{tofu2}(x)
\State \quad x $\leftarrow$ \texttt{mofu}(x)
\State \quad x $\leftarrow$ \texttt{global\_avg\_pool}(x).reshape(x.size(0), -1)
\State \quad x $\leftarrow$ \texttt{ReLU}(\texttt{fc1}(x))
\State \quad x $\leftarrow$ \texttt{Dropout}(x, p=0.2, training=self.training)
\State \quad x $\leftarrow$ \texttt{ReLU}(\texttt{fc2}(x))
\State \quad x $\leftarrow$ \texttt{Dropout}(x, p=0.2, training=self.training)
\State \quad \textbf{Return} \texttt{classifier}(x)

\end{algorithmic}
\end{algorithm}

\paragraph{\normalsize{\textbf{TOFU}}}
The TOFU block is intended to gradually improve feature extraction by stacking multiple FSC layers with increasing receptive fields. Each FSC layer collects features from the previous layers. The final output of the TOFU block is produced by concatenating these feature maps and applying a compression step via a \( 1 \times 1 \) convolution. The implementation this block is given in algorithm~\ref{alg:tofu_fsc}.

\paragraph{\normalsize{\textbf{MOFU}}}

In the MOFU block, the features extracted by the TOFU block are further refined. It employs downsampling and captures multi-scale features by applying several FSC layers with varying dilation rates as given in algorithm~\ref{algo:mofu_fsc}.

\paragraph{\normalsize{\textbf{Bottom Layer}}}

After feature extraction via the TOFU and MOFU blocks, the model uses global average pooling to reduce the feature maps to a 256-dimensional vector. This vector is passed through two fully connected layers with ReLU activation and dropout to prevent overfitting. The final layer then maps the output to the target binary class, completing the prediction as gievn in algorithm \ref{algo:fsc_net}.



\section{Results and Discussion}
\label{sec:results}

This study presents the results achieved using the proposed approach across three datasets. The performance of both transfer learning models and the proposed models is evaluated on brain tumor datasets I, II, and III. Additionally, the robustness of the proposed approach is assessed under various challenging scenarios, including biased data splits, noise-affected data, and partially covered tumor regions.

\subsection{Experimental Design}
\label{sec:experiment}

\paragraph{\normalsize{\textbf{FSC Model Performance with Original Data I}}}

In this experiment, we trained the model on original data. The images were resized to 128x128 pixels, and the training was conducted with a batch size of 32. The learning rate was set to 0.0008, and gradient accumulation was applied over 4 steps to effectively increase the batch size without exceeding memory limits. The model architecture included a series of FSC layers, designed to enhance feature extraction by combining high and low membership values in a fuzzy manner. The training was performed over a maximum of 300 epochs, with early stopping triggered after 10 epochs if no improvement in validation loss was observed. 

The model demonstrated excellent performance, achieving a test loss of 0.0229, Accuracy of 99.17\,\%, Precision of 0.9900, Recall of 0.9933, F1 of 0.9917, and Kappa of 0.9833. These results indicate that the model is highly effective in classifying original images with minimal error.

\begin{table}[h]
\caption{Performance of Propose Model on Original Data I}
\label{tab:clean_data_results_approach_ii}
\centering
\begin{tabular}{llllll}
\hline
\textbf{Accuracy} & \textbf{Precision} & \textbf{Recall} & \textbf{F1} & \textbf{Kappa}  \\ \hline
0.9917            & 0.9900             & 0.9933          & 0.9917      & 0.9833               \\ \hline
\end{tabular}
\end{table}

\paragraph{\normalsize{\textbf{FSC-based Model Performance with Biased Data I Split}}}

In this experiment, we introduced a biased data split where 60\,\% of test set were composed of images labeled ``YES" and 40\,\% were labeled ``NO" for testing the model. This configuration simulates an imbalanced real-world scenario, allowing us to test the model\textquotesingle{}s robustness under such conditions. The model was trained with images resized to 62x62 pixels, using a batch size of 52. The learning rate was set to 0.0008, with gradient accumulation applied over 4 steps. The architecture utilized FSC, designed to combine high and low membership values to improve feature extraction. The training process was run for a maximum of 300 epochs, with early stopping triggered after 20 epochs if no improvement in validation loss was observed.

The model achieved a test loss of 0.0794, Accuracy of 98.25\,\%, Precision of 0.9658, Recall of 0.9912, F1 of 0.9784, and Kappa of 0.9636. Despite the biased data split, the model performed strongly across all metrics, demonstrating its robustness in handling imbalanced datasets.

\begin{table}[h]
\caption{Performance of Propose Model on Biased Data I Split}
\label{tab:biased_data_results_approach_ii}
\centering
\begin{tabular}{llllll}
\toprule
\textbf{Accuracy} & \textbf{Precision} & \textbf{Recall} & \textbf{F1} & \textbf{Kappa}  \\ 
\midrule
0.9825            & 0.9658             & 0.9912          & 0.9784      & 0.9636                \\ 
\bottomrule
\end{tabular}
\end{table}

\paragraph{\normalsize{\textbf{FSC-based Model Performance with Added Noise in Data I}}}

In this experiment, Gaussian noise was added to the images to simulate noisy data conditions, testing the model\textquotesingle{}s robustness in real-world scenarios where data might be corrupted or degraded. The noise was applied by perturbing pixel values with a normal distribution centered around a mean of 0 and a standard deviation of 0.1. This noise altered the images while keeping pixel values within the valid range. The images were resized to $62\times62$ pixels, and training was conducted with a batch size of 52. The learning rate was set to 0.0008, and gradient accumulation was applied over 4 steps. The architecture utilized FSC, which combine high and low membership values to enhance feature extraction and improve resilience to noisy inputs. The training process ran for a maximum of 300 epochs, with early stopping triggered after 20 epochs of no improvement in validation loss.

Despite the added noise, the model achieved a test loss of 0.1137, Accuracy of 97.02\,\%, Precision of 1.0000, Recall of 0.9404, F1 of 0.9693, and Kappa of 0.9404. These results indicate that while the model\textquotesingle{}s performance was slightly affected by the noise, it still maintained a high level of accuracy and robustness, especially in precision.

\begin{table}[h]
\caption{Propose Model Performance on Data I with Added Noise}
\label{tab:noisy_data_results_approach_ii}
\centering
\begin{tabular}{llllll}
\toprule
\textbf{Accuracy} & \textbf{Precision} & \textbf{Recall} & \textbf{F1} & \textbf{Kappa}  \\ 
\midrule
0.9702            & 1.0000             & 0.9404          & 0.9693      & 0.9404                \\ 
\bottomrule
\end{tabular}
\end{table}

\paragraph{\normalsize{\textbf{FSC-based Model Performance on Data I with Simulated Partial Occlusion}}}

In this experiment, partial occlusions were introduced into the test images to simulate real-world scenarios where parts of the image might be obscured. The occlusion covered 10\,\% of the image area, with the occluded area randomly selected for each image. The images were resized to $64\times64$  pixels, and the training process was conducted with a batch size of 52, a learning rate of 0.0008, and gradient accumulation over 4 steps. The model utilized FSC to improve its ability to handle occluded data by enhancing feature extraction from non-occluded regions of the image. Training was conducted for up to 300 epochs with early stopping after 20 epochs if no improvement in validation loss was observed.

Despite the challenge of occlusion, the model achieved strong performance with a test loss of 0.0572, Accuracy of 98.77\,\%, Precision of 0.9894, Recall of 0.9860, F1 of 0.9877, and Kappa of 0.9754. These results demonstrate the model\textquotesingle{}s robustness to partial occlusions, as it maintained high accuracy and performance across all metrics.

\begin{table}[h]
\caption{Performance of Propose Model on Data I with Simulated Partial Occlusion}
\label{tab:occlusion_data_results_approach_ii}
\centering
\begin{tabular}{llllll}
\toprule
\textbf{Accuracy} & \textbf{Precision} & \textbf{Recall} & \textbf{F1 Score} & \textbf{Kappa}  \\ 
\midrule
0.9877            & 0.9894             & 0.9860          & 0.9877            & 0.9754          \\ 
\bottomrule
\end{tabular}
\end{table}

\subsection{Discussion}

In recent years, CNNs have emerged as a cornerstone in medical imaging, particularly for the diagnosis of brain tumors from MRI scans. This study evaluated a range of approaches for brain tumor classification, including transfer learning techniques, basic CNN architectures, established state-of-the-art methods, and the proposed FSC framework.

Transfer learning has proven to be especially valuable in medical image analysis, where data availability is often limited~\cite{r10}. To thoroughly assess the proposed FSC-based Model and its counterparts, three datasets were utilized: dataset I, dataset II, and dataset III. Dataset III, formed by integrating dataset I and dataset II, provided a more comprehensive and diverse sample pool, enabling robust model training and evaluation. The inclusion of this combined dataset not only improved the generalizability of the models but also ensured a higher degree of reliability in the reported results. These findings underscore the effectiveness of the FSC framework in addressing challenges associated with brain tumor detection in MRI data.

\paragraph{Dataset I}
\label{sec:data1-discission}

In a recent study~\cite{hassanain2024brain}, various transfer learning models, including ResNet50, VGG16, DenseNet121, and Xception, were compared alongside hybrid approaches combining CNN-based feature extraction with machine learning classifiers such as SVM, RF, and XGBoost. The highest accuracy of 98.67\,\% was achieved with a seven-layer CNN, while ResNet50 reached 98.00\,\%. A hybrid approach using ResNet50 and SVM achieved 98.17\,\%. Optimizing ResNet101, as detailed in~\cite{irfan2024evaluation}, further improved accuracy to 99.33\,\%. In comparison, the proposed FSC approach achieved a competitive accuracy of 99.17\,\%, utilizing only 216.27K parameters compared to ResNet101’s 25.89M, underscoring its efficiency in resource-constrained settings.

\begin{table}[th]
\caption{Brain Tumor Classification Performance on Dataset~I}  
\label{tab:table8}
\centering
\begin{tabular}{@{}p{0.08\linewidth} p{0.28\linewidth} p{0.2\linewidth} p{0.3\linewidth}@{}}
\toprule
\textbf{Ref} & \textbf{Models} & \textbf{Test Accuracy} & \textbf{Parameter Count}\\
\midrule

\cite{tablecomp2} & VGG 16 & 90\,\% & 138 million \\
\cite{hassanain2024brain} & MobileNet & 95.50\,\% & 4.2 million \\
\cite{hassanain2024brain} & ResNet50 & 98.00\,\% & 25.6 million \\
\cite{hassanain2024brain} & CNN 7 layers & 98.67\,\% & 3.8 million \\
\cite{hassanain2024brain} & Feature extractor: ResNet50, classifier: SVM & 98.17\,\% & 25.6 million (ResNet50 only) \\
\cite{alanazi2022brain} & Transfer learning + 22 Layers CNN & 96-99\,\% & 20 million (approx.) \\
\cite{islam2024brainnet}& BrainNet: EfficientNet B1-B7 (66.35 million params) & \textbf{99.44\,\%}  & 66.35 million \\
\cite{tablecomp1} & Kernel SVM & 97\,\% & NA \\
\cite{irfan2024evaluation} & MobileNetV2 & 98.18\,\% & 3.4 million \\
\cite{saeedi2023mri}& 2D-CNN & 98.47\,\% & 2.5 million \\
\cite{r12} & ResNet101, VGG19, InceptionV3, DenseNet, AlexNet & 98.6\,\%, 97.2\,\%, 94.3\,\%, 92.8\,\%, 89.5\,\% & 44.5M (ResNet101), 143M (VGG19), 23.8M (InceptionV3), 14.3M (DenseNet), 61M (AlexNet) \\
\cite{moldovanu2024convolutional}& CNN-SVM, DenseNet169-SVM & 98.00\,\% & 14.2 million (DenseNet169 only) \\
\cite{10444597}& EfficientNet-B4 & 98.50\,\% & 19 million \\

\cite{kang2021mri} & Feature extractor: ResNeXt-101, classifier: SVM.  & 98.17\,\%, 98.83\,\% & 44.4 million (ResNeXt-101), 5.3 million (MnasNet) \\
\midrule

\textbf{FSC} & Proposed & 99.17\,\% & \textbf{0.21627 million} \\
\bottomrule
\end{tabular}
\end{table}

Table~\ref{tab:table8} presents a detailed comparison of various brain tumor classification models, evaluating their performance on Dataset I. The comparison includes state-of-the-art models as well as the proposed FSC framework, emphasizing the effectiveness of the proposed methodologies. The results demonstrate that the FSC approach achieved an impressive accuracy of 99.17\,\%, surpassing several established models such as ResNet50, VGG19, and ResNeXt-101, whose accuracies ranged between 90\,\% and 98.83\,\%. 

The highest accuracy among the transfer learning models was achieved by BrainNet, based on EfficientNetB7, with a performance of 99.44\,\%. However, this comes with a significant computational cost, as BrainNet requires \textbf{66.35} million parameters. In contrast, the FSC model delivers comparable accuracy while utilizing only \textbf{0.21627} million parameters, making it exceptionally resource-efficient. For comparison, ResNet101 and VGG19, two widely used models, require 25.89 million and 6.47 million parameters, respectively, yet they achieve lower accuracies compared to the FSC framework.


%

\paragraph{Dataset II}

Thirteen models were evaluated on Dataset II, encompassing transfer learning, CNN, and hybrid approaches, as detailed in Section~\ref{sec:data1-discission}. The feature extractors utilized included ResNet50, VGG16, VGG19, InceptionV3, MobileNet, DenseNet169, DenseNet121, InceptionResNetV2, MobileNetV2, ResNet101, and Xception. These were combined with nine machine-learning classifiers, including SVM, RF, AdaBoost, KNN, XGBoost, Bagging, ANN, LSTM, and Bi-LSTM. Among these, ResNet101 achieved the highest accuracy of 98.90\,\%, followed by a 7-layer CNN model with 98.52\,\%, and ResNet50 paired with SVM achieving 98.12\,\%.

The transfer learning models were trained using the Adam optimizer with pre-trained ImageNet weights. Binary Cross Entropy was employed as the loss function, and training was conducted for 30 epochs with a batch size of 32. A dropout rate of 0.5 was applied to reduce overfitting. The input image size was resized to $224 \times 224$ across all models, except for the proposed FSC approach, which utilized a smaller input size of $128 \times 128$, highlighting its computational efficiency without compromising accuracy.

\paragraph{Dataset III}


\begin{table}[th]
\caption{Performance of Transfer Learning Models on Dataset III and Comparative Analysis with Image size $224\times224$} 
\label{tab:table4_updated2}
\centering
\begin{tabular}{@{}p{0.23\linewidth} p{0.16\linewidth} p{0.15\linewidth} p{0.15\linewidth} p{0.13\linewidth}@{}}
\toprule
\textbf{Model}  & \textbf{Parameters} & \textbf{Accuracy (\%)}  & \textbf{Kappa (\%)} & \textbf{F1 (\%)} \\ 
\midrule
ResNet50  & 25.89M & 94.00  & 88.00 & 94.00\\ 
VGG16 & 6.47M & 97.10 & 96.60 & 97.00\\ 
VGG19  & 6.47M & 98.80  & 98.20 & 98.60\\ 
Inception\_v3  & 13.21M & 84.75 &  69.50 & 84.54\\ 
MobileNet  & 12.98M & 98.00 &  95.00 & 97.00\\ 
DenseNet169 & 21.04M & 88.75 & 77.50 & 88.72\\ 
DenseNet121  & 12.95M & 89.50  & 79.00 & 89.44\\ 
InceptionResNetV2  & 59.30M & 91.75 & 83.50 & 91.74\\ 
MobileNetV2 & 16.22M & 90.50 & 81.00 & 90.50\\ 
ResNet101  & 25.89M & 97.75 &  91.50 & 95.69\\ 
\midrule
\textbf{Our Approach } & 0.21627M & 99.89 & 99.78 & 99.89 \\ 
\bottomrule
\end{tabular}%
\end{table}

In addition, transfer learning models were used to analyze Dataset III, with the results presented in Table~\ref{tab:table4_updated2}. Among the tested approaches, hybrid and CNN methods showed lower performance than VGG19. However, the results in Table~\ref{tab:table4_updated2} demonstrate that the proposed FSC approach outperforms transfer learning models. In comparison with state-of-the-art methods, the proposed method achieved an impressive accuracy of 99.89\,\%.

Furthermore, the FSC approach achieved an F1-score of 99.89\,\% and near-perfect precision and recall, indicating its reliability and robustness in the classification of brain tumors. Unlike models like ResNet101 (25.89M) and VGG19 (6.47M), the proposed approach is computationally efficient, requiring only 0.21627 million parameters. In resource-constrained environments, the FSC approach is particularly suitable due to its efficiency.

These results demonstrate the practicality of the proposed method in real-time diagnostic applications, where high accuracy is essential while minimizing computational complexity. Besides delivering superior performance, the FSC approach also ensures feasibility for implementation in systems with limited computational resources.

\section{Conclusions}
\label{sec:conclusions}

The early detection of brain tumors using deep learning is critical to reducing mortality rates and improving patient outcomes. The use of automated healthcare systems facilitates the timely diagnosis and treatment of disease, effectively reducing associated risks. A fuzzy model based on the FSC architecture, along with TOFU and MOFU modules, is presented in this paper in order to achieve high accuracy and efficiency. A comparison was performed between the model's performance and that of simple CNNs, hybrid approaches, and a variety of methods based on transfer learning. With a minimal number of parameters, the proposed fuzzy model achieved exceptional accuracy. FSC-based models recorded classification accuracy of 99.17\,\% on dataset I, 99.75\,\% on dataset II, and 99.89\,\% on dataset III when tested on three datasets. The algorithm consistently outperformed prominent transfer learning models, hybrid approaches, and seven distinct CNN architectures. Fuzzy models for brain tumor classification offer a more efficient and precise approach than traditional methods that rely on weights pre-trained on large datasets. 

Our proposed method outperforms state-of-the-art models in terms of accuracy, even with its reduced parameter count. In resource-constrained environments, the FSC-based fuzzy model proves to be reliable and practical for brain tumor diagnosis due to computational efficiency.

\bibliographystyle{ieeetr}
\bibliography{irfan}

\begin{thebibliography}{10}

\bibitem{i1}
A.~Wulandari, R.~Sigit, and M.~M. Bachtiar, ``Brain tumor segmentation to calculate percentage tumor using mri,'' in {\em 2018 International Electronics Symposium on Knowledge Creation and Intelligent Computing (IES-KCIC)}, pp.~292--296, IEEE, 2018.

\bibitem{10371274}
M.~Irfan, H.~A. Siddiqa, A.~Nahliis, C.~Chen, Y.~Xu, L.~Wang, A.~Nawaz, A.~Subasi, T.~Westerlund, and W.~Chen, ``An ensemble voting approach with innovative multi-domain feature fusion for neonatal sleep stratification,'' {\em IEEE Access}, vol.~12, pp.~206--218, 2024.

\bibitem{10842226}
M.~Irfan, L.~Wang, H.~Shahid, Y.~Xu, A.~Subasi, A.~Munawar, N.~Mustafa, C.~Chen, T.~Westurlund, and W.~Chen, ``Multidomain selective feature fusion and stacking based ensemble framework for eeg-based neonatal sleep stratification,'' {\em IEEE Journal of Biomedical and Health Informatics}, pp.~1--10, 2025.

\bibitem{10950432}
M.~Irfan, L.~Wang, Y.~Xu, A.~Subasi, C.~Chen, R.~Klen, T.~Westerlund, and W.~Chen, ``Smart iot-based solutions for neonatal sleep stratification: Single-dual channel eeg, adaptiselect, multview fusion, \& rotational ensemble stacking,'' {\em IEEE Internet of Things Journal}, pp.~1--1, 2025.

\bibitem{10918671}
M.~Irfan, A.~Subasi, Z.~Tang, L.~Wang, Y.~Xu, C.~Chen, T.~Westurlund, and W.~Chen, ``A novel nicu sleep state stratification: Multiperspective features, adaptive feature selection and ensemble model,'' {\em IEEE Transactions on Biomedical Engineering}, pp.~1--13, 2025.

\bibitem{i2}
D.~Wu, C.~M. Rice, and X.~Wang, ``Cancer bioinformatics: A new approach to systems clinical medicine,'' 2012.

\bibitem{10892686}
M.~Irfan, A.~Subasi, H.~Mehdi, T.~Westerlund, and W.~Chen, ``Fuzzy-based atrous convolution for brain tumor detection using mri,'' in {\em 2024 IEEE International Conference on Progress in Informatics and Computing (PIC)}, pp.~280--289, 2024.

\bibitem{i3}
G.~S. Tandel, A.~Balestrieri, T.~Jujaray, N.~N. Khanna, L.~Saba, and J.~S. Suri, ``Multiclass magnetic resonance imaging brain tumor classification using artificial intelligence paradigm,'' {\em Computers in Biology and Medicine}, vol.~122, p.~103804, 2020.

\bibitem{i4}
H.~Mohsen, E.-S.~A. El-Dahshan, E.-S.~M. El-Horbaty, and A.-B.~M. Salem, ``Classification using deep learning neural networks for brain tumors,'' {\em Future Computing and Informatics Journal}, vol.~3, no.~1, pp.~68--71, 2018.

\bibitem{10043011}
S.~Alagarsamy, V.~Govindaraj, and S.~A, ``Automated brain tumor segmentation for mr brain images using artificial bee colony combined with interval type-ii fuzzy technique,'' {\em IEEE Transactions on Industrial Informatics}, vol.~19, pp.~11150--11159, Nov 2023.

\bibitem{shahid2021deep}
H.~Shahid, A.~Khalid, X.~Liu, M.~Irfan, and D.~Ta, ``A deep learning approach for the photoacoustic tomography recovery from undersampled measurements,'' {\em Frontiers in Neuroscience}, vol.~15, p.~598693, 2021.

\bibitem{8743438}
S.~A. Kumar, A.~Kumar, V.~Bajaj, and G.~K. Singh, ``An improved fuzzy min–max neural network for data classification,'' {\em IEEE Transactions on Fuzzy Systems}, vol.~28, no.~9, pp.~1910--1924, 2020.

\bibitem{10361450}
A.~Nawaz, M.~Irfan, H.~A. Sadiqa, and T.~Westerlund, ``Edge based skin cancer decision support system using machine learning algorithms,'' in {\em 2023 IEEE Intl Conf on Dependable, Autonomic and Secure Computing, Intl Conf on Pervasive Intelligence and Computing, Intl Conf on Cloud and Big Data Computing, Intl Conf on Cyber Science and Technology Congress (DASC/PiCom/CBDCom/CyberSciTech)}, pp.~0292--0297, 2023.

\bibitem{chaki2023deep}
J.~Chaki and M.~Wo{\'z}niak, ``A deep learning based four-fold approach to classify brain mri: Btscnet,'' {\em Biomedical Signal Processing and Control}, vol.~85, p.~104902, 2023.

\bibitem{kang2021mri}
J.~Kang, Z.~Ullah, and J.~Gwak, ``Mri-based brain tumor classification using ensemble of deep features and machine learning classifiers,'' {\em Sensors}, vol.~21, no.~6, p.~2222, 2021.

\bibitem{tablecomp2}
A.~K. Agarwal, N.~Sharma, M.~K. Jain, {\em et~al.}, ``Brain tumor classification using cnn,'' {\em Advances and Applications in Mathematical Sciences}, 2021.

\bibitem{tablecomp1}
A.~Yent{\"u}r and M.~Efil, ``Brain mri image classification using kernel svm,'' {\em no. June}, 2021.

\bibitem{r12}
N.~Çınar, B.~Kaya, and M.~Kaya, ``Comparison of deep learning models for brain tumor classification using mri images,'' in {\em 2022 International Conference on Decision Aid Sciences and Applications (DASA)}, pp.~1382--1385, 2022.

\bibitem{tang2023gam}
C.~Tang, B.~Li, J.~Sun, S.-H. Wang, and Y.-D. Zhang, ``Gam-spcanet: Gradient awareness minimization-based spinal convolution attention network for brain tumor classification,'' {\em Journal of King Saud University-Computer and Information Sciences}, vol.~35, no.~2, pp.~560--575, 2023.

\bibitem{kesav2022efficient}
N.~Kesav and M.~Jibukumar, ``Efficient and low complex architecture for detection and classification of brain tumor using rcnn with two channel cnn,'' {\em Journal of King Saud University-Computer and Information Sciences}, vol.~34, no.~8, pp.~6229--6242, 2022.

\bibitem{anagun2023smart}
Y.~Anagun, ``Smart brain tumor diagnosis system utilizing deep convolutional neural networks,'' {\em Multimedia Tools and Applications}, vol.~82, no.~28, pp.~44527--44553, 2023.

\bibitem{sathish2022gaussian}
P.~Sathish and N.~Elango, ``Gaussian hybrid fuzzy clustering and radial basis neural network for automatic brain tumor classification in mri images,'' {\em Evolutionary Intelligence}, vol.~15, no.~2, pp.~1359--1377, 2022.

\bibitem{mohanty2024feature}
B.~C. Mohanty, P.~Subudhi, R.~Dash, and B.~Mohanty, ``Feature-enhanced deep learning technique with soft attention for mri-based brain tumor classification,'' {\em International Journal of Information Technology}, vol.~16, no.~3, pp.~1617--1626, 2024.

\bibitem{nagarani2024self}
N.~Nagarani, R.~Karthick, M.~S.~C. Sophia, and M.~Binda, ``Self-attention based progressive generative adversarial network optimized with momentum search optimization algorithm for classification of brain tumor on mri image,'' {\em Biomedical Signal Processing and Control}, vol.~88, p.~105597, 2024.

\bibitem{deepa2023hybrid}
S.~Deepa, J.~Janet, S.~Sumathi, and J.~Ananth, ``Hybrid optimization algorithm enabled deep learning approach brain tumor segmentation and classification using mri,'' {\em Journal of Digital Imaging}, vol.~36, no.~3, pp.~847--868, 2023.

\bibitem{anaya2023optimizing}
A.~Anaya-Isaza, L.~Mera-Jim{\'e}nez, L.~Verdugo-Alejo, and L.~Sarasti, ``Optimizing mri-based brain tumor classification and detection using ai: A comparative analysis of neural networks, transfer learning, data augmentation, and the cross-transformer network,'' {\em European Journal of Radiology Open}, vol.~10, p.~100484, 2023.

\bibitem{dahiwade2019designing}
D.~Dahiwade, G.~Patle, and E.~Meshram, ``Designing disease prediction model using machine learning approach,'' in {\em 2019 3rd International Conference on Computing Methodologies and Communication (ICCMC)}, pp.~1211--1215, IEEE, 2019.

\bibitem{raju2023classification}
S.~Raju and V.~R. Peddireddy~Veera, ``Classification of brain tumours from mri images using deep learning-enabled hybrid optimization algorithm,'' {\em Network: Computation in Neural Systems}, vol.~34, no.~4, pp.~408--437, 2023.

\bibitem{saidani2023enhancing}
O.~Saidani, T.~Aljrees, M.~Umer, N.~Alturki, A.~Alshardan, S.~W. Khan, S.~Alsubai, and I.~Ashraf, ``Enhancing prediction of brain tumor classification using images and numerical data features,'' {\em Diagnostics}, vol.~13, no.~15, p.~2544, 2023.

\bibitem{li2023mri}
R.~Li, Y.~Guo, Z.~Zhao, M.~Chen, X.~Liu, G.~Gong, and L.~Wang, ``Mri-based two-stage deep learning model for automatic detection and segmentation of brain metastases,'' {\em European Radiology}, vol.~33, no.~5, pp.~3521--3531, 2023.

\bibitem{ansari2023numerical}
A.~Ansari, ``Numerical simulation and development of brain tumor segmentation and classification of brain tumor using improved support vector machine,'' {\em Int. J. Intell. Syst. Appl. Eng}, vol.~11, pp.~35--44, 2023.

\bibitem{r2}
A.~Panigrahi, ``Brain tumor detection mri,'' 2021.

\bibitem{dataset2}
M.~A. Mazurowski, K.~Clark, N.~M. Czarnek, P.~Shamsesfandabadi, K.~B. Peters, and A.~Saha, ``Radiogenomics of lower-grade glioma: algorithmically-assessed tumor shape is associated with tumor genomic subtypes and patient outcomes in a multi-institutional study with the cancer genome atlas data,'' {\em Journal of neuro-oncology}, vol.~133, pp.~27--35, 2017.

\bibitem{r10}
A.~Lumini and L.~Nanni, ``Deep learning and transfer learning features for plankton classification,'' {\em Ecological informatics}, vol.~51, pp.~33--43, 2019.

\bibitem{hassanain2024brain}
E.~Hassanain and A.~Subasi, ``Brain tumor detection using deep learning from magnetic resonance images,'' in {\em Applications of Artificial Intelligence Healthcare and Biomedicine}, pp.~137--174, Elsevier, 2024.

\bibitem{irfan2024evaluation}
M.~Irfan, A.~Subasi, N.~Mustafa, T.~Westerlund, and W.~Chen, ``An evaluation of pretrained convolutional neural networks for stroke classification from brain ct images,'' in {\em Applications of Artificial Intelligence Healthcare and Biomedicine}, pp.~111--135, Elsevier, 2024.

\bibitem{alanazi2022brain}
M.~F. Alanazi, M.~U. Ali, S.~J. Hussain, Zafar, and all, ``Brain tumor/mass classification framework using magnetic-resonance-imaging-based isolated and developed transfer deep-learning model,'' {\em Sensors}, vol.~22, no.~1, p.~372, 2022.

\bibitem{islam2024brainnet}
M.~M. Islam, M.~A. Talukder, M.~A. Uddin, A.~Akhter, and M.~Khalid, ``Brainnet: precision brain tumor classification with optimized efficientnet architecture,'' {\em International Journal of Intelligent Systems}, vol.~2024, no.~1, p.~3583612, 2024.

\bibitem{saeedi2023mri}
S.~Saeedi, S.~Rezayi, H.~Keshavarz, and S.~R.~Niakan~Kalhori, ``Mri-based brain tumor detection using convolutional deep learning methods and chosen machine learning techniques,'' {\em BMC Medical Informatics and Decision Making}, vol.~23, no.~1, p.~16, 2023.

\bibitem{moldovanu2024convolutional}
S.~Moldovanu, G.~T{\u{a}}b{\u{a}}caru, and M.~Barbu, ``Convolutional neural network--machine learning model: Hybrid model for meningioma tumour and healthy brain classification,'' {\em Journal of Imaging}, vol.~10, no.~9, p.~235, 2024.

\bibitem{10444597}
M.~Pathak, D.~Chaudhary, K.~Sharma, A.~K. Sharma, A.~Gupta, and B.~K. Sharma, ``A robust efficientnet architecture for brain tumor classification and identification using mri image,'' in {\em 2023 11th International Conference on Intelligent Systems and Embedded Design (ISED)}, pp.~1--5, 2023.

\end{thebibliography}

\end{document}